\documentclass[aps,prb,twocolumn,superscriptaddress,nofootinbib]{revtex4-2}

\usepackage{amsmath,amssymb,physics,bm}
\usepackage[T1]{fontenc}
\usepackage{lmodern}
\usepackage{xspace}
\usepackage{xcolor}
\usepackage{graphicx}
\usepackage{epsfig}
\usepackage{times}
\usepackage[normalem]{ulem}
\usepackage[colorlinks,linkcolor=blue,citecolor=blue,anchorcolor=blue,urlcolor=blue]{hyperref}

\newcommand{\bl}{\begin{aligned}}
\newcommand{\el}{\end{aligned}}
\def\be{\begin{equation}}
\def\ee{\end{equation}}

\def\bi{\begin{itemize}}
\def\ei{\end{itemize}}
\def\bn{\begin{enumerate}}
\def\en{\end{enumerate}}
\def\bea{\begin{eqnarray}}
\def\eea{\end{eqnarray}}

\def\ba{\begin{array}}
\def\ea{\end{array}}
\def\bd{\begin{displaymath}}
\def\ed{\end{displaymath}}

\begin{document}

\title{Reply to Comment on ``Scaling and universality at noisy quench dynamical quantum phase transitions''}

\author{S. Ansari}
\affiliation{Department of Engineering Sciences and Physics, Buein Zahra Technical University, Buein Zahra 34518-66391, Iran}

\author{R.~Jafari}
\email{raadmehr.jafari@gmail.com}
\affiliation{Department of Physics, Institute for Advanced Studies in Basic Sciences (IASBS), Zanjan 45137-66731, Iran}
\affiliation{School of Quantum Physics and Matter Science, Institute for Research in Fundamental Sciences (IPM), Tehran 19395-5531, Iran}

\author{A. Akbari}
\affiliation{Beijing Institute of Mathematical Sciences and Applications (BIMSA), Huairou District, Beijing 101408, China}

\author{M. Abdi}
\affiliation{School of Physics and Astronomy, Shanghai Jiao Tong University, Shanghai 200240, China}

\date{\today}

\begin{abstract}
The Comment by J.~Sirker~\cite{Sirker:2025:arXiv} raises an important issue
concerning dynamical quantum phase transitions (DQPTs) in noisy and mixed-state
dynamics, namely that the extension of the Loschmidt echo from pure to mixed
states is not unique and different extensions preserve different physical
properties.
The Comment examines a noise-averaged mixed-state fidelity and shows that DQPTs cannot occur for any nonzero noise
when the return rate is defined through the Uhlmann-Bures fidelity of the noise-averaged density matrix.
This conclusion is valid for the mixed-state fidelity observable discussed in the Comment and is consistent with prior studies~\cite{Jafari:2024:PRB:b,Jafari:2025:arXiv:a}.
Our article~\cite{Ansari:2025:PRB} investigated a different operationally defined quantity: the logarithm of the Loschmidt echo obtained by first determining
the noise-averaged excitation probabilities generated during the noisy ramp 
and then performing a coherent post-ramp evolution of a pure state constructed
from these noise-averaged transition probabilities.
As emphasized explicitly in our original publication, this observable is defined through an operational assumption and
is not the same quantity as the mixed-state fidelity.
The nonanalyticities reported in Ref.~\cite{Ansari:2025:PRB} therefore concern this two-stage operational protocol and should not be identified with zeros of the Uhlmann-Bures fidelity.
There is therefore no direct contradiction between the theorem established for the Uhlmann-Bures return rate and the conclusions obtained for the different operational protocol studied in Ref.~\cite{Ansari:2025:PRB}.
\end{abstract}

\maketitle

\paragraph*{Agreement with the Comment's formalism.}
We first emphasize the point on which there is no disagreement.
The Comment provides a clear classification of several noise-averaging
procedures for Loschmidt-type observables. This taxonomy is fully consistent
with earlier analyses, including
Refs.~\cite{Jafari:2024:PRB:b,Jafari:2025:arXiv:a}. The fidelity definition
studied in the Comment corresponds to one specific member of this taxonomy,
namely the return rate constructed from the Uhlmann-Bures
fidelity between noise-averaged mixed states. Within that definition the Comment correctly
proves that dynamical quantum phase transitions cannot occur for any finite
noise.

The central issue is therefore not the validity of this theorem, but its domain of applicability.
Indeed, extending the Loschmidt echo from pure to mixed states is not unique: different constructions preserve different physical properties and therefore probe different aspects of nonequilibrium dynamics.
Our PRB article \cite{Ansari:2025:PRB} did not employ this mixed-state fidelity. 
Instead, it introduced explicitly an operational assumption: the post-ramp state is represented by a pure state whose
populations match the noise-averaged excitation probabilities obtained from
the master equation. This operational choice was explicitly stated in the
original publication and corresponds to a different Loschmidt-type protocol
from the Uhlmann-Bures construction considered in the Comment. Since the observable considered in our work is not
equivalent to the mixed-state fidelity used in the Comment, the no-DQPT
conclusions proven there do not apply to the operational protocol analyzed in
our PRB.
\\

\paragraph*{Standard approaches to noise averaging.}
For clarity we summarize the two noise-averaging procedures that are most
commonly used in the study of DQPTs and that also appear in the taxonomy of
the Comment.

\begin{enumerate}

\item \textit{Averaging over single noise realizations.}
The system is initialized in the ground state at $t_i$, $|g_k(t_i)\rangle$,
and the noisy ramp is applied up to $t_f \rightarrow 0^{-}$. For each
individual realization of the noise the Loschmidt echo is
\bea
L^{(S)}_k(t)=|\langle g_k(t_f)\vert g_k(t)\rangle|^2 .
\eea
Averaging is then performed over realizations at fixed time, producing
$\langle L_k(t)\rangle$. As shown in
Ref.~\cite{Jafari:2024:PRB:b}, DQPTs occur in each realization but at
different critical times, so the averaging procedure removes the
nonanalyticities and leads to no DQPT.

\item \textit{Averaging over the noisy ramp only.}
Here the averaging is performed over the noise that acts during the interval
$t_i$ to $t_f$ by solving the exact noise master equation, which yields a
mixed state $\rho_k(t_f)$ ~\cite{Jafari:2025:PRR,Jafari:2025:PRB:a,Jafari:2025:arXiv:a}. The natural generalization of the Loschmidt echo
within a fidelity-based mixed-state formulation
is then the Uhlmann-Bures fidelity. As demonstrated in
Refs.~\cite{Jafari:2024:PRB:b,Jafari:2025:arXiv:a}, this
fidelity-based return rate has no zeros for any finite noise intensity.
Hence DQPTs are absent within this definition.

\end{enumerate}
These two procedures correspond to two branches of the classification
discussed in the Comment. They are mathematically well-defined and physically useful, but they do not exhaust all possible operational definitions of Loschmidt-type observables.
The observable employed in our PRB does not belong
to either of these fidelity-based categories, since it is based on a different operational
assumption about the post-ramp state.
\\

\paragraph*{What our results in Ref.~\cite{Ansari:2025:PRB} actually studied.}
In our original article noise acts only during the ramp.
Solving the exact
noise master equation yields a mixed state $\rho_k(t_f)$ at the end of the
ramp. 

Our work therefore introduced an explicit operational assumption motivated by
coherent post-ramp dynamics: rather than evaluating the Loschmidt return rate directly from the mixed state,
we considered a population-preserving pure state whose excitation probability
matches that obtained from the noise-averaged master equation.

 The state used at $t = 0$ was
\begin{equation}
|\psi_k(0)\rangle
  = \sqrt{1 - p_k}\,|g_k^{f}\rangle
    + \sqrt{p_k}\,|e_k^{f}\rangle ,
\end{equation}
where $|g_k^{f}\rangle$ and $|e_k^{f}\rangle$ are the post-ramp eigenstates
and $p_k$ is the noise-averaged excitation probability obtained from the master equation.
This construction was not introduced as the optimal pure-state
approximation to $\rho_k(t_f)$ in the Uhlmann-Bures fidelity.
Rather, it defines an operational protocol in which the
noise-averaged populations obtained during the ramp are used as
input for a subsequent coherent evolution. The protocol is
experimentally motivated because mode-resolved excitation
probabilities are directly measurable in engineered two-level
systems and quantum simulators
\cite{Ai:2021:PRA,Yang2019,Ma2018,APL:2020}.
\\

\paragraph*{Dynamical free energy.}
Within this operational protocol the return rate is determined by evolving
the population-matched pure state under the noiseless post-ramp Hamiltonian
$H_f$. The resulting dynamical free energy takes the form
\begin{equation}
  g(t)
  = -\frac{1}{2\pi}\int_{0}^{\pi} dk\,
    \ln\!\Big[
    1 - 4\,p_k(1 - p_k)\,
    \sin^{2}\!\big(\varepsilon_k^{f} t\big)
    \Big].
  \label{eq:g}
\end{equation}
Equation~(\ref{eq:g}) is precisely the return rate analyzed
in Ref.~\cite{Ansari:2025:PRB}. Unlike the Uhlmann-Bures
return rate, it is constructed from the operational protocol
defined above.
Related interferometric and two-band simulation schemes provide experimentally relevant ways to access Loschmidt-type amplitudes in coherent few-level settings~\cite{Yang2019,Ma2018,APL:2020}.
Although the pure state $|\psi_k(0)\rangle$ is not identical to the mixed
state $\rho_k(t_f)$, the excitation probability $p_k$ encodes information
about both diagonal and off-diagonal components of $\rho_k(t_f)$ once it is
transformed into the adiabatic basis. This point is important because the Comment argues that a protocol based on excitation probabilities is blind to coherence information.
We show below, through explicit
Landau-Zener analysis, that $p_k$ is generally a function of both diagonal and off-diagonal elements of the density matrix in the diabatic basis.
This supports
the internal consistency of the observable evaluated in our paper and explains
why nonanalytic behavior can emerge within this operational framework.
\\[3mm]

%
\begin{figure}
\centerline{
\includegraphics[width=0.995\linewidth]{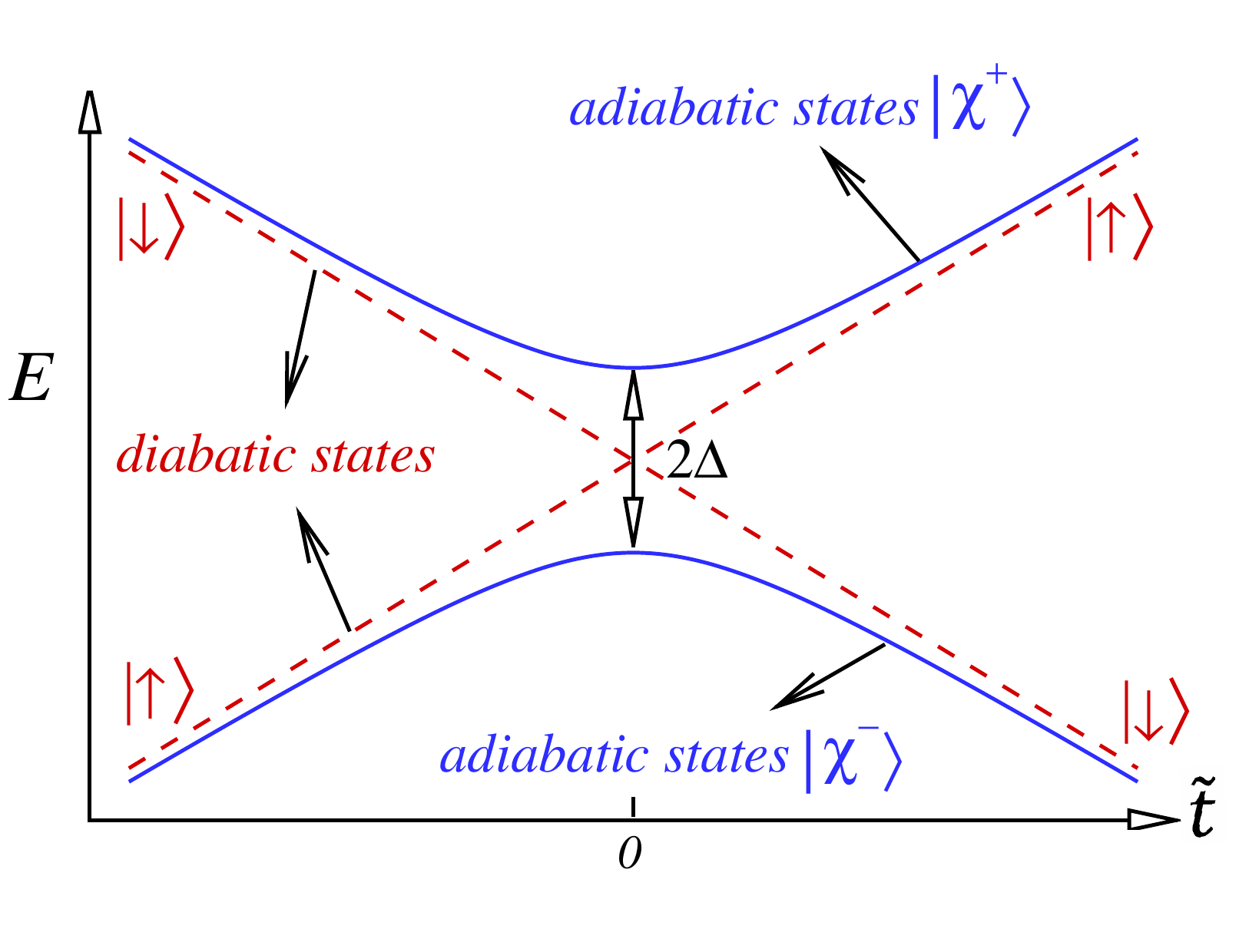}}
\caption{Adiabatic (solid) and diabatic (dashed) energies for the
standard Landau-Zener problem.}
\label{fig1}
\end{figure}
%

\paragraph*{Landau-Zener analysis.}

The revised Comment argues that replacing the mixed state by a
pure state specified through $p_k$ neglects coherence
information. To address this point we analyze the generic
Landau-Zener problem, where one can explicitly show that
$p_k$ depends on both diagonal and off-diagonal elements of
the density matrix after transformation to the adiabatic basis.

We consider the standard time-dependent two-level Hamiltonian
\begin{equation}
\label{eq1}
H(t) =
\begin{pmatrix}
  h(t) & \Delta \\
  \Delta & -h(t)
\end{pmatrix},
\end{equation}
where $h(t)$ is a slowly varying control field and $\Delta$ is the minimum
gap. The instantaneous eigenstates of $H(t)$ form the adiabatic basis.
They are given by
\begin{align}
\label{eq:eigenstates}
|\chi^{-}(t)\rangle
  &= \alpha(t)\,|\uparrow\rangle
     + \beta(t)\,|\downarrow\rangle
     = \begin{pmatrix} \alpha(t) \\[1mm] \beta(t) \end{pmatrix},
\\[2mm]
|\chi^{+}(t)\rangle
  &= -\beta(t)\,|\uparrow\rangle
     + \alpha(t)\,|\downarrow\rangle
     = \begin{pmatrix} -\beta(t) \\[1mm] \alpha(t) \end{pmatrix},
\end{align}
with corresponding instantaneous energies
\begin{equation}
\label{eq:spectrum}
\varepsilon^{\pm}(t)
  = \pm\,\varepsilon(t),
\qquad
\varepsilon(t)=\sqrt{h^{2}(t)+\Delta^{2}}.
\end{equation}
Here the coefficients $\alpha(t) \!= \!(h(t)\!-\!\varepsilon(t))/N(t)$, and \,$\beta(t)
\!=\! \Delta/N(t)$,
encode the rotation between
the diabatic and adiabatic bases, and their explicit forms depend only on
$h(t)$, $\Delta$, and the normalization convention $N(t)=\sqrt{2\varepsilon(t)(\varepsilon(t)-h(t))}$. This parametrization
makes it possible to express $\rho_k(t_f)$ in either basis and to show how the
measured excitation probability
\bea
\label{excitation-probability}
p_k = \langle \chi^{+}(t_f)|\rho_k(t_f)|\chi^{+}(t_f)\rangle
\eea
combines contributions from both the diagonal and off-diagonal matrix elements
of $\rho_k(t_f)$ in the diabatic basis. 
\\

%
%
%
The unitary transformation under which the Hamiltonian is diagonal at fixed $t'$ is given as
%
\bea
H^{(d)}(t')=U^{\dag}(t')H(t')U(t')
=\left(
  \begin{array}{cc}
    -\varepsilon(t') & 0 \\
    0 & \varepsilon(t') \\
  \end{array}
\right),
\eea
%
with 
%
\begin{equation}
U(t')=\!
\left(
  \begin{array}{cc}
    \alpha(t') & -\beta(t') \\
    \beta(t') & \alpha(t') \\
  \end{array}
\right);
\quad
U^{\dag}(t')=\!
\left(
  \begin{array}{cc}
    \alpha(t') & \beta(t') \\
    -\beta(t') & \alpha(t') \\
  \end{array}
\right).  
\end{equation}
%
If the system is initially prepared in the ground state \( |\chi^{-}(t_i)\rangle \) at time \( t_i \), then to obtain the transition probability to the upper level \( |\chi^{+}(t')\rangle \) at time \( t' \), one must solve the von Neumann equation or the noise averaged master equation.
%
\begin{equation}
\label{eq:master}
  \dot \rho_k(t)= -i\big[H_k^{(0)}(t),\rho_k(t)\big]
  - \frac{\xi^2}{2}\big[H_1,\big[H_1,\rho_k(t)\big]\big].
\end{equation}
%
Since the Hamiltonian is given in the diabatic basis ($|\uparrow\rangle$, $|\downarrow\rangle$), the density matrix $\rho(t)$ obtained from Eq.~\eqref{eq:master} is expressed in that basis. Therefore, the diagonal elements of $\rho(t)$ are diabatic populations, not adiabatic excitation probabilities.
To obtain the excitation probability at a fixed time $t'$ in the adiabatic basis, the density matrix should be transformed to the adiabatic basis $|\chi^{\pm}(t')\rangle$ using 
the unitary transformation, i.e., 
\begin{widetext}
%
\bea
\label{eq:rhoadiabatic}
\rho^{(d)}(t')=U^{\dag}(t')\rho(t')U(t')=
\left(
  \begin{array}{cc}
    \alpha(t') & \beta(t') \\
    -\beta(t') & \alpha(t') \\
  \end{array}
\right)
\left(
  \begin{array}{cc}
    \rho_{11}(t') & \rho_{12}(t') \\
    \rho_{21}(t') & \rho_{22}(t') \\
  \end{array}
\right)
\left(
  \begin{array}{cc}
    \alpha(t') & -\beta(t') \\
    \beta(t') & \alpha(t') \\
  \end{array}
\right),
\eea
%
which gives
%
\begin{equation}
\label{eq:rhodmatrix}
\rho^{(d)}(t')=
\left(
  \begin{array}{cc}
    \alpha ^2 \rho_{11}+\alpha  \beta  (\rho_{12}+\rho_{21})+\beta ^2 \rho_{22} ~~&~~ \alpha ^2 \rho_{12}+\alpha  \beta  (\rho_{22}-\rho_{11})-\beta ^2 \rho_{21} \\
    \alpha ^2 \rho_{21}+\alpha  \beta  (\rho_{22}-\rho_{11})-\beta ^2 \rho_{12} ~~&~~  \alpha ^2 \rho_{22}-\alpha  \beta  (\rho_{12}+\rho_{21})+\beta ^2 \rho_{11}\\
  \end{array}
\right),
\end{equation}
%
%
\end{widetext}
where $\rho^{(d)}_{11}(t')$ and  $\rho^{(d)}_{22}(t')$ are the probabilities of occupying the lower and upper adiabatic levels at $t'$, respectively. 
Equivalently,
%
\begin{equation}
\label{eq:rhoadiabaticelements}
\rho^{(d)}(t')=
\left(
  \begin{array}{cc}
   \langle\chi^{-}(t')|\rho(t')|\chi^{-}(t')\rangle  & \langle\chi^{-}(t')|\rho(t')|\chi^{+}(t')\rangle \\
   \langle\chi^{+}(t')|\rho(t')|\chi^{-}(t')\rangle  & \langle\chi^{+}(t')|\rho(t')|\chi^{+}(t')\rangle \\
  \end{array}
\right), 
\end{equation}
%
which yields
\begin{equation}
\bl
\label{eq:rho11d}
\rho^{(d)}_{11}(t')
&=\langle\chi^{-}(t')|\rho(t')|\chi^{-}(t')\rangle  
\\
&=\alpha ^2 \rho_{11}
+\alpha\beta(\rho_{12}+\rho_{21})
+\beta ^2 \rho_{22},
\\
\rho^{(d)}_{12}(t')
&=\langle\chi^{-}(t')|\rho(t')|\chi^{+}(t')\rangle 
\\
&=\alpha ^2 \rho_{12}
+\alpha\beta(\rho_{22}-\rho_{11})
-\beta ^2 \rho_{21},
\\
\rho^{(d)}_{21}(t')
&=\langle\chi^{+}(t')|\rho(t')|\chi^{-}(t')\rangle 
\\
&=\alpha ^2 \rho_{21}
+\alpha\beta(\rho_{22}-\rho_{11})
-\beta ^2 \rho_{12},
\\
\rho^{(d)}_{22}(t')
&=\langle\chi^{+}(t')|\rho(t')|\chi^{+}(t')\rangle 
\\
&=\alpha ^2 \rho_{22}
-\alpha\beta(\rho_{12}+\rho_{21})
+\beta ^2 \rho_{11}.
\el
\end{equation}
Equations~\eqref{eq:rhodmatrix} and \eqref{eq:rho11d} show explicitly that the adiabatic excitation probability
\begin{equation}
\bl
\label{eq:pkcoh}
p_k=
&\rho^{(d)}_{22}(t_f)
\\
=
&\alpha ^2 \rho_{22}(t_f)
-\alpha\beta\!\left[\rho_{12}(t_f)+\rho_{21}(t_f)\right]
+\beta ^2 \rho_{11}(t_f).
\el
\end{equation}

contains the off-diagonal elements of the density matrix in the diabatic basis, except in special limits where the basis rotation is trivial. These expressions show explicitly that the excitation probability entering our protocol generally depends on both diagonal and off-diagonal elements of the density matrix in the diabatic basis. Consequently, the operational protocol should not, in general, be interpreted as retaining only classical population information.
In the presence of noise the time evolution of the density matrix is given by the exact noise master equation which has been used in Refs.~\cite{Anirban2016,Sadeghizade2025,Gao2017}.
For more details about the difference between transition probability in adiabatic and diabatic state please see Eq.~(4) and Eq.~(7) in Ref.~\cite{Vitanov1999} {\color{black}and also Refs. \cite{Rahmani2016,Singh2021,Ai2021,Griffin2012}} 
\\

\paragraph*{Position of our observable within the taxonomy.}
In our protocol noise acts only during the ramp, while the post-ramp evolution
is fully coherent under the same Hamiltonian $H_f$ for all realizations. The
only quantity carried over from the noisy stage is the excitation probability, Eq.~(\ref{excitation-probability}),
which is obtained from the noise-averaged master equation. Inserting this
population into Eq.~\eqref{eq:g} determines the return rate. A nonanalyticity
arises whenever there exists a momentum $k^\ast$ with $p_{k^\ast}=1/2$, which
is the condition highlighted in our paper.
This condition is a property of the operational return rate in Eq.~\eqref{eq:g}; it is not claimed to be a zero of the Uhlmann-Bures fidelity.

The operational assumption underlying this construction was
explicitly stated in Ref.~\cite{Ansari:2025:PRB}. Therefore,
the no-DQPT theorem established for the Uhlmann-Bures fidelity
does not apply to the observable investigated in our work.
\\
%
%
\paragraph*{Operational interpretation.}
Our protocol has a direct experimental interpretation consisting of two
sequential steps:
\begin{enumerate}
\item Apply the noisy ramp, solve the master equation, and measure the
mode-resolved excitation probabilities $p_k$ at $t=0$.
\item Prepare for each $k$ a pure state with noise-averaged populations $(1-p_k,p_k)$ and let
it evolve coherently under the fixed Hamiltonian $H_f$.
\end{enumerate}
The return rate in our paper is precisely the interferometric signal generated
by this second, fully coherent stage.

As noted in the revised Comment, this protocol may be viewed
as an interferometric protocol rather than as a measurement
of the Uhlmann-Bures fidelity.
%
%
%
%
\paragraph*{Conclusion.}
The Comment's results~\cite{Sirker:2025:arXiv}, which correctly demonstrate
the absence of DQPTs when using the mixed-state fidelity, are in agreement with
previous studies~\cite{Jafari:2024:PRB:b,Jafari:2025:arXiv:a}. Our claim in
Ref.~\cite{Ansari:2025:PRB}, however, concerns a different and experimentally
motivated observable based on population-preserving pure states, a distinction
that was clearly stated in the original publication. Within that operationally
realizable protocol the return rate can exhibit nonanalyticities. 
Moreover, although not required for the arguments presented here,
the physical picture underlying the dynamical phase structure reported
in Ref.~\cite{Ansari:2025:PRB} is further supported by an independent
theoretical approach developed in our subsequent
work~\cite{Jafari_PRBLetter2026}, which reaches the same qualitative
conclusions using a different formulation and connected to the DQPTs using the reduced fidelity \cite{Parez2026a, Parez2026b}. 
Therefore, while the Comment establishes an important no-go theorem for the
Uhlmann-Bures mixed-state return rate, it does not invalidate the
conclusions obtained for the operational two-stage protocol studied
in Ref.~\cite{Ansari:2025:PRB}.


%

\end{document}